\documentclass{aastex631}
\usepackage{amsmath}
\usepackage{comment} % 
\usepackage{subfigure}
\usepackage{booktabs}
\usepackage{multirow} % Add this to use the multirow command
\usepackage{makecell}
\usepackage{threeparttable} % For adding notes under the table
\usepackage{CJK}
\usepackage{makecell}
\usepackage{threeparttable}
\usepackage{float}         % 
\usepackage{placeins}

\begin{document}
\begin{CJK*}{UTF8}{gbsn}
%\linenumbers

\title{Identifying Red Supergiants in the Local Group Using JWST Photometry. I. NGC 6822, Sextans A, NGC 300, WLM, and IC 1613}

\author[0009-0003-3321-6393]{Zhi-wen Li (李志文)}
\affiliation{Institute for Frontiers in Astronomy and Astrophysics,
            Beijing Normal University,  Beijing 102206, China}
\affiliation{School of Physics and Astronomy,
               Beijing Normal University,
               Beijing 100875, China}

\author[0000-0001-8247-4936]{Ming Yang (杨明)}
\affiliation{Key Laboratory of Space Astronomy and Technology,
             National Astronomical Observatories,
             Chinese Academy of Sciences,
             Beijing 100101, China}

\author[0000-0003-3168-2617]{Biwei Jiang (姜碧沩)}
\affiliation{Institute for Frontiers in Astronomy and Astrophysics,
            Beijing Normal University,  Beijing 102206, China}
\affiliation{School of Physics and Astronomy,
               Beijing Normal University,
               Beijing 100875, China}

\author[0000-0003-1218-8699]{Yi Ren (任逸)}
\affiliation{Department of Astronomy,
            College of Physics and Electronic Engineering,
            Qilu Normal University,
            Jinan 250200, China}

\correspondingauthor{Ming Yang, Biwei Jiang}
\email{myang@nao.cas.cn, bjiang@bnu.edu.cn}




%% automatically renumbers them.

\begin{abstract}
Red supergiants (RSGs) are crucial for studying the properties and evolution of massive stars. It is representative to conduct a census of RSGs across the Local Group, which spans a broad metallicity range. However, identifying RSGs in distant and metal-poor galaxies remains challenging mainly due to contamination of foreground dwarfs and observational limitations. In this work, we perform PSF photometry on publicly released JWST/NIRCam images of five Local Group galaxies: NGC 6822, Sextans~A, NGC 300, WLM, and IC 1613 using the DOLPHOT NIRCam module. We find an optimal color-color diagram (CCD) for metal-poor environments, that is F115W $-$ F200W versus F356W $-$ F444W, which clearly separates RSGs from foreground dwarfs. By using the CCD, we identify 208, 135, and 22 RSG candidates in NGC 6822, Sextans A, and NGC 300, respectively, free from contamination by foreground dwarfs and oxygen-rich asymptotic giant branch stars (O-AGBs). In addition, 40 and 14 RSG candidates are directly selected on the CMD in WLM and IC 1613, respectively. Compared with previous works, the number of RSG candidates within the same luminosity range and sky region increases significantly, demonstrating the advantages of JWST in constructing a more complete RSG sample in the Local Group thanks to its high spatial resolution and photometric quality. In addition, catalogs of O-AGBs and carbon-rich AGBs (C-AGBs) are provided as by-products.
\end{abstract}


\section{Introduction}
Red supergiants (RSGs) are Population~I massive stars in the core helium-burning stage and are the most luminous cool stars. They are the progenitors of Type II-P supernovae. Typically, RSGs have initial masses of about $6.5$--$30~M_\odot$, effective temperatures ($T_{\mathrm{eff}}$) of $\sim3500$--$4500$~K, radii of $\sim100$--$1000~R_\odot$, luminosities of $\sim4\times10^{3}$--$4\times10^{5}~L_\odot$, and surface gravities ($\log g$) of $\sim 0$~\citep{1979ApJ...232..409H,2005ApJ...628..973L,2013EAS....60...31E,2013NewAR..57...14M,2017ApJ...847..112D, 2024ApJ...965..106Y}. They are relatively young, with ages of $\sim8$--$35$~Myr~\citep{2019A&A...631A..95B}, spending about $10^{6}$--$10^{7}$~years on the main sequence. After exhausting hydrogen in their cores, they evolve to the RSG phase on a very short timescale of about $10^{3}$--$10^{5}$~years~\citep{2017ars..book.....L}. The RSG phase itself lasts only $\sim10^{5}$--$10^{6}$~years~\citep{2013A&A...558A.103G}, during which RSGs occupy the upper-right region of the optical to near-infrared color-magnitude diagram (CMD), forming a slightly sloped and well-defined red branch.

RSGs play a crucial role in tracing recent star formation~\citep{2011MNRAS.411..235E,2019IAUS..344...77H,2024ApJ...966...25R}, addressing the mass loss mystery of massive stars~\citep{2001ApJ...551.1073H, 2010ApJ...717L..62Y, 2011A&A...526A.156M, 2016MNRAS.463.1269B, 2023A&A...676A..84Y, 2024ApJS..275...33W, 2024A&A...686A..88A,2025arXiv250305876A,2025Galax..13...72V, 2025A&A...697A.167Z}, calibrating the period-luminosity relation~\citep{2006MNRAS.372.1721K, 2011ApJ...727...53Y, 2012ApJ...754...35Y, 2018ApJ...859...73S, 2019MNRAS.487.4832C, 2019ApJS..241...35R, 2024IAUS..376..292J}, investigating the binary fraction among massive stars~\citep{2020ApJ...900..118N,2021ApJ...908...87N,2022MNRAS.513.5847P,2025A&A...700A..36P, 2025MNRAS.539.1220D, 2025ApJ...988...60D}, studying the scaling relations between granulation and stellar parameters~\citep{2020ApJ...898...24R, 2024ApJ...969...81Z}, and exploring RSGs as cosmic abundance probes~\citep{2015ApJ...806...21D,2015ApJ...803...14P}. 
As the cornerstone of these studies, conducting a complete census of RSGs across a broad range of metallicities is essential. Hence, we require a comprehensive census across the Local Group, where metallicities, i.e., [Fe/H], range from approximately $+0.3$ to $-2.5$~\citep{2012AJ....144....4M}.

Identifying RSGs in the Milky Way is a challenging task due to the significant and inhomogenous extinction in the Galactic plane. Even so, several studies search for RSGs in the Galaxy using spectroscopic~
\citep{2025A&A...694A.152Z}, granulation variability~\citep{2025MNRAS.538..101Z}, and CMD-based~\citep{2024MNRAS.529.3630H} methods. Consequently, the primary focus of RSG identification shifts to nearby galaxies, where the member stars have the same distance. The main difficulty in identifying extragalactic RSGs lies in distinguishing them from Galactic foreground dwarfs, as they share nearly identical apparent colors and magnitudes, which can cause severe contamination on the CMD. 
Several methods have been proposed to separate RSGs from foreground dwarfs, including radial-velocity and spectroscopic classification \citep{2016ApJ...826..224M} and machine-learning techniques \citep{2022A&A...666A.122M,2024ApJ...965..106Y}. However, these approaches are often limited by data availability or photometric quality, which prevents the identification of faint RSGs and thus hinders the construction of a complete sample.
Another commonly used approach is selecting RSGs through a combination of multiple CMDs \citep{2020ApJ...892...91H,2022ApJ...933..197T}, but it still suffers from foreground contamination.
The more effective and widely adopted methods are the color-color diagram (CCD) method~\citep{1998ApJ...501..153M,2021A&A...647A.167Y,2021ApJ...907...18R,2021ApJ...923..232R,2025ApJ...979..208L} and the Gaia astrometric method~\citep{2019A&A...629A..91Y,2021A&A...646A.141Y,2022AJ....163...70D,2025ApJ...979..208L}.

A large and relatively complete sample of RSGs has been constructed for four Local Group galaxies: SMC, LMC, M31, and M33 \citep{2019A&A...629A..91Y, 2021A&A...646A.141Y, 2021ApJ...907...18R,2021ApJ...923..232R, 2020ApJ...900..118N, 2020ApJ...889...44N, 2021AJ....161...79M}. These samples are constructed using CCD criteria as well as Gaia astrometric information, which effectively remove contamination from Galactic foreground dwarfs.
For M31 and M33, the $B-V$/$V-R$, $r-z$/$z-H$, and $J-H$/$H-K$ CCDs are adopted in separating RSGs from Galactic foreground dwarfs, as there is a clear bifurcation between the two populations on these diagrams in the metal-rich environments ([Fe/H] $\sim$ 0.3 for M31 and $\sim$ 0.0 for M33; \citealt{2018MNRAS.478.5379D, 1995RMxAC...3..133E, 1997ApJ...489...63G, 2025ApJ...979..208L}).
The photometry-based CCD method has no distance limitation compared with astrometric method. However, it becomes inoperative for metal-poor galaxies, such as the LMC and SMC ([Fe/H] $\sim$ $-0.5$ and $-1.0$ respectively, \citealt{2012AJ....144....4M}), because RSGs at low metallicities blend into the foreground dwarf branch on these CCDs \citep{2025ApJ...979..208L}. Fortunately, for the LMC and SMC, the foreground dwarfs are efficiently removed by using Gaia astrometric data, due to their relatively nearby distances ($\sim$ 50 kpc for the LMC and $\sim$ 60 kpc for the SMC).
For the moderately distant ($m-M=23.40$, $d = 479$ kpc; \citealt{2012MNRAS.421.2998F}) and metal-poor galaxy NGC 6822 ([Fe/H] $\sim$ $-1.0$; \citealt{2012AJ....144....4M}), where Gaia data is very incomplete and RSGs blend into the dwarf branch on the CCD, \citet{2025ApJ...979..208L} combine Gaia astrometry with CCD methods to construct a more complete RSG sample.
However, identifying RSGs in other galaxies of the Local Group remains challenging, as these galaxies are generally distant and metal-poor. The Gaia astrometric method is limited by distance, and the CCD method is hindered by metallicity. Therefore, developing a new CCD for metal-poor environments would be a significant advancement in distinguishing RSGs from Galactic foreground dwarfs in distant and metal-poor galaxies.

Nowadays, with the new data available from the James Webb Space Telescope (JWST; \citealt{2023PASP..135f8001G}), which is equipped with new filters, new opportunities arise to develop new CCDs. Furthermore, space-based observations offer high spatial resolution and photometry quality, greatly avoiding the misidentification of blended adjacent sources as a single object~\citep{2025ApJ...988...60D}. The improved photometric precision also helps disentangle the overlap between RSGs and oxygen-rich asymptotic giant branch (O-AGB) stars on the CMD.
\citet{2024MNRAS.531..183N} unveil previously unseen stellar populations in NGC 6822 using the JWST CMD especially in the faint end.
\citet{2024ApJ...973..120B} investigate various CCD combinations based on the JWST F090W, F150W, F250M, and F430M bands to distinguish C-AGB and O-AGB stars.
In this study, by combining the JWST F090W, F115W, F200W, F356W (or F335M), and F444W (or F430M) bands, we look for an optimal CCD for metal-poor environments.

This paper is organized as follows. Section~\ref{Sect.obs_and_pho} introduces the JWST observations and photometry. Section~\ref{Sect.data} presents the data reduction.
Section~\ref{Sect.Method} describes the method used to remove foreground dwarfs and identify RSGs on the CMD and CCD.
Section~\ref{Sect.Application} introduces the application of the method to additional galaxies.
Section~\ref{Sect.results} presents the results, Section~\ref{Sect.comparison} compares them with previous studies, and Section~\ref{Sect.conclusion} is a summary.

\section{Observation and Photometry}\label{Sect.obs_and_pho}
\subsection{Observation}
JWST has observed approximately twenty galaxies in the Local Group, with observations for about half of them publicly available to date. Considering the size and coverage of the observed fields and the availability of photometric bands, we select NGC 6822, Sextans A, NGC 300, WLM, and IC 1613. All of these galaxies, except IC 1613, contain a key band near 4~$\mu$m (F444W or F435M), which serves as a new diagnostic band to distinguish RSGs from foreground dwarfs in the combined CCDs

\begin{figure}
    \centerline{\includegraphics[width=0.7\linewidth]{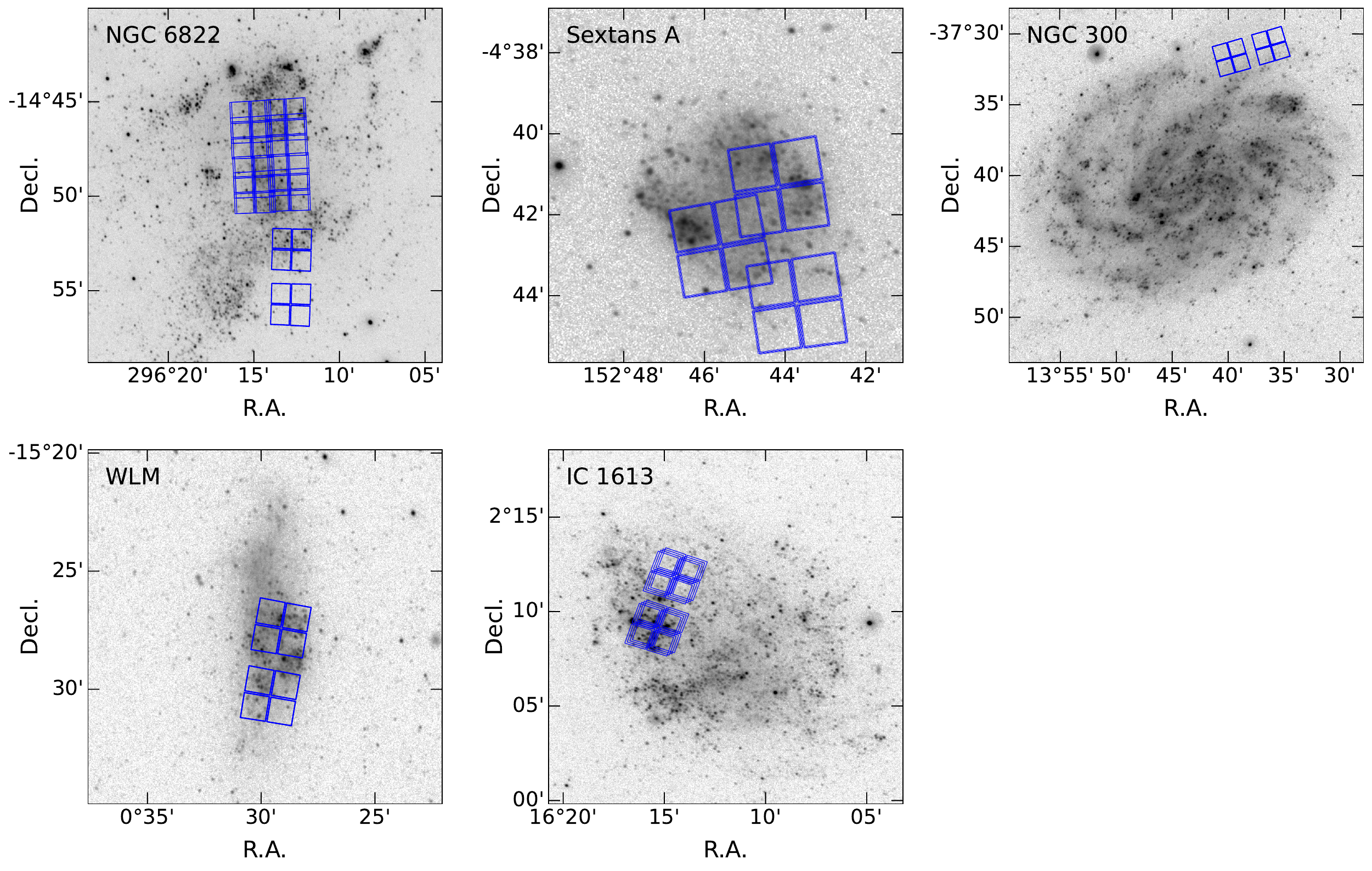}}
    \caption{The observed fields of view of NGC 6822, Sextans A, NGC 300, WLM, and IC 1613. The blue frames mark the JWST fields used for photometry in this study, overlaid on GALEX ultraviolet images.}
    \label{Figure.Observation}
\end{figure}

The data used in this study are from various JWST programs: ID 1234 (PI: Margaret Meixner) for NGC 6822, ID 1619 (PI: Martha L. Boyer) for Sextans A, ID 1638 (PI: Kristen B. W. McQuinn) for NGC 300, ID 1334 (PI: Daniel R. Weisz) for WLM, and ID 2391 (PI: Julia Roman-Duval) for IC 1613. 
All of the data presented in this article are obtained from the Mikulski Archive for Space Telescopes (MAST\footnote{\url{https://mast.stsci.edu/portal/Mashup/Clients/Mast/Portal.html}}) at the Space Telescope Science Institute. The specific observations analyzed can be accessed as follows. \dataset[doi:10.17909/zfgd-yj09]{https://doi.org/10.17909/zfgd-yj09} (NIRCam data for NGC 6822),
\dataset[doi:10.17909/mpx2-7m41]{https://doi.org/10.17909/mpx2-7m41} (NIRCam data for Sextans A),
\dataset[doi:10.17909/86ek-t361]{https://doi.org/10.17909/86ek-t361} (NIRCam data for NGC 300),
\dataset[doi:10.17909/kd96-v198]{https://doi.org/10.17909/kd96-v198} (NIRCam data for WLM), and
\dataset[doi:10.17909/ac71-c267]{https://doi.org/10.17909/ac71-c267} (NIRCam data for IC 1613).
The observation fields for these galaxies are shown in Figure~\ref{Figure.Observation} by blue frames, and detailed observation information is listed in Table~\ref{Table.jwst_observation}.

\subsection{Photometry}
For NGC 6822 (Observtn 9, denoting the observation number in the MAST archive), Sextans A, NGC 300, WLM, and IC 1613, a total of four dithered exposures are used for photometry. For the other NGC 6822 dataset (Observtn 6 and 10), only four out of the twelve available dithered exposures (exposure numbers 1, 4, 7, and 10) are used, because they already cover the intended field of view, equivalent to that of the full set of twelve exposures, and ensure a photometric depth consistent with the other observations. Besides, four exposures already provide sufficient photometric precision for studying bright sources. The dithered areas are shown as blue frames in Figure~\ref{Figure.Observation}. 
All photometric science images are downloaded from MAST. The science images are the Stage 2 files (cal.fits), which are calibrated single-exposure images produced by the JWST pipeline.

We perform point-spread function (PSF) photometry on these images using the DOLPHOT NIRCam module \citep{2000PASP..112.1383D, 2016ascl.soft08013D, 2024ApJS..271...47W}. DOLPHOT is a straightforward and efficient photometry package, and a detailed description of its implementation for NIRCam data is provided by \citet{2024ApJS..271...47W}. 
In brief, the photometric procedure consists of three steps.
First, it masks bad pixels based on the Data Quality Flags in the image headers, and calculates the sky background.
Second, it aligns all science images to a common reference frame using a selected Stage 3 image (i2d.fits) as the reference, ensuring that the coordinates from different fields are registered consistently in the final catalog. The F150W image is used as the reference because of its better source detection performance \citep{2023ApJ...956L..18R} for all galaxies except NGC~6822 for which the F200W image is used to achieve a similar effect.
Third, it detects stellar sources and performs simultaneous PSF photometry across different fields and bands.

The JWST NIRCam (0.6--5.0~$\mu$m) is divided into the short-wavelength (SW; 0.6--2.3~$\mu$m) and long-wavelength (LW; 2.4--5.0~$\mu$m) channels. 
Although photometry can be performed on the SW and LW images simultaneously, photometry is first conducted on the SW images alone to obtain the SW-band measurements, and then on the combined SW+LW images to obtain the LW-band measurements, following the procedure in \citet{2023ApJ...956L..18R}. 
This approach is adopted because the lower resolution of the LW images may lead to inaccurate detections and aperture corrections, introducing a photometric deviation of approximately 2~mmag in the SW bands \citep{2023ApJ...956L..18R}. 
Using the same photometric parameters as in \citet{2024ApJS..271...47W}, photometry for all five galaxies is carried out following the above procedure. 
After the separate photometry steps, the SW- and LW-band measurements are combined through crossmatching, with the PSF FWHM of the reference image adopted as the matching radius. 
The photometric results for the SW and LW bands and the used crossmatch radius are listed in Table~\ref{Table.photometry_result}.

\section{Data Reduction}\label{Sect.data}
After obtaining the photometric data, we perform quality control following the criteria of \citet{2024ApJS..271...47W}: 
(1) $Sharpness_{\mathrm{SW}}^2 \leq 0.01$ to remove spurious sources, 
(2) $Crowding_{\mathrm{SW}} \leq 0.5$ to ensure photometric reliability, 
(3) $PhotometryFlag_{\mathrm{SW}} \leq 3$ to control the photometric quality and to exclude sources that are severely affected by saturation, and 
(4) $ObjectType \leq 2$ to retain point sources. 

\begin{figure}
    \centerline{\includegraphics[width=0.75\linewidth]{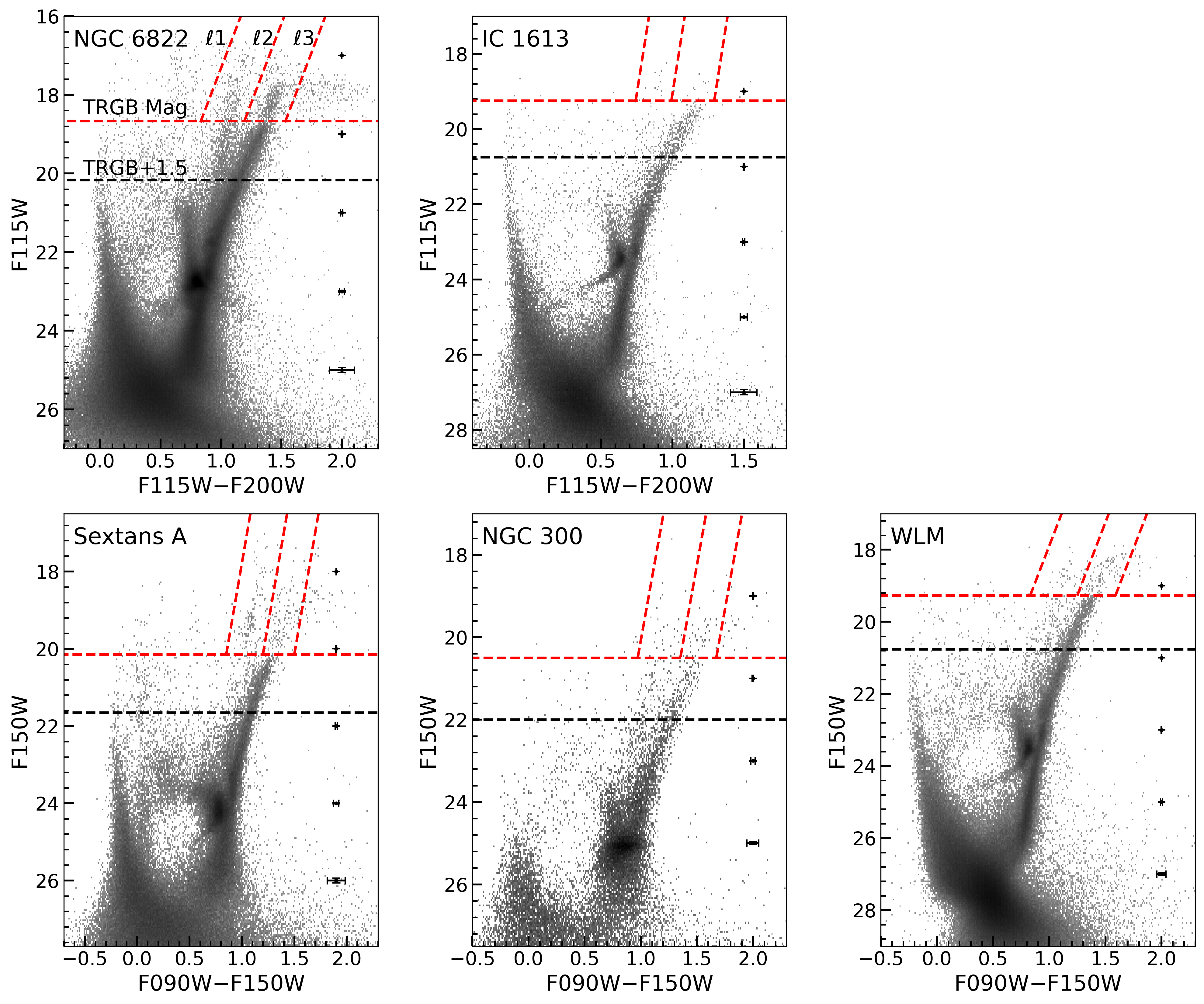}}
    \caption{Various CMDs based on the photometric results for NGC 6822, IC 1613, Sextans A, NGC 300, and WLM. In each CMD, the red dashed line denotes the magnitude of TRGB, while the sloped dashed lines $l_1$, $l_2$, and $l_3$ represent the boundaries of RSGs, O-AGBs, and C-AGBs. Sources brighter than the black dashed line are analyzed in this study.}
    \label{Figure.Cat_CMD}
\end{figure}

The numbers of sources satisfying these criteria are listed in Table~\ref{Table.photometry_result}. 
To remove duplicate detections, the data are further self-crossmatched using the corresponding search radius, and one detection is retained for each matched group. The resulting statistics are summarized in Table~\ref{Table.photometry_result}. 
The final dataset (Table~\ref{Table.photometry_catalog}) is publicly available for retrieval, and the CMDs of the galaxies are shown in Figure~\ref{Figure.Cat_CMD}. Structures of the main sequence, red giant branch (RGB), red clump, and horizontal branch are clearly visible on the CMDs, as well as the RSG branch and AGB population at the bright end.

\section{Distinguish RSGs and Foreground Dwarfs on JWST CMD and CCD}
\label{Sect.Method}
To understand the distribution of RSGs and their overlap with foreground dwarfs on the JWST CMD, we first define the regions occupied by these two populations. The typical metal-poor galaxy NGC 6822 ([Fe/H]~$\sim -1.0$; \citealt{2012AJ....144....4M, 2003PASP..115..635D, 2022Univ....8..465R, 2011ASPC..445..409S, 2020ApJ...892...91H}) is adopted as a test case. 
Then, we introduce a newly constructed JWST CCD for metal-poor environments with the F444W band. The RSG candidates are first selected on the CMD by the defined RSG region, then the foreground dwarfs are subsequently removed by using the CCD.

\subsection{RSGs on the CMD}
We use the PARSEC stellar evolution model to define the approximate locations of different stellar populations on the JWST CMD, which facilitates their subsequent identification on the CCD. The CMD 3.8 web interface\footnote{\url{https://stev.oapd.inaf.it/cgi-bin/cmd_3.8}} is used to generate the PARSEC isochrones. The input parameters are as follows. 
The evolutionary tracks are based on PARSEC version~1.2S combined with 
COLIBRI~S\_37, COLIBRI~S\_35, and COLIBRI~PR16, 
with $n_{\mathrm{inTPC}} = 10$ and $\eta_{\mathrm{Reimers}} = 0.2$. 
The photometric system adopts the JWST NIRCam wide and very wide filters. 
Bolometric corrections are taken from the YBC library with the new Vega calibration. 
The circumstellar dust composition is assumed to be 
100\% silicate for M-type stars, and a mixture of 85\% amorphous carbon and 15\% SiC for C-type stars. 
The interstellar extinction is set to $A_V = 0$. 
Long-period variability is included following \citet{2021MNRAS.500.1575T}. 
The initial mass function (IMF) follows the canonical two-part power law 
and is corrected for unresolved binaries \citep{2002Sci...295...82K}. 
The model grid covers $\log t$ from 6.0 to 10.5 with a step of 0.05, 
and adopts a metallicity of [Fe/H]~$=-1.0$, appropriate for NGC 6822.

\begin{figure}
    \centerline{\includegraphics[width=0.95\linewidth]{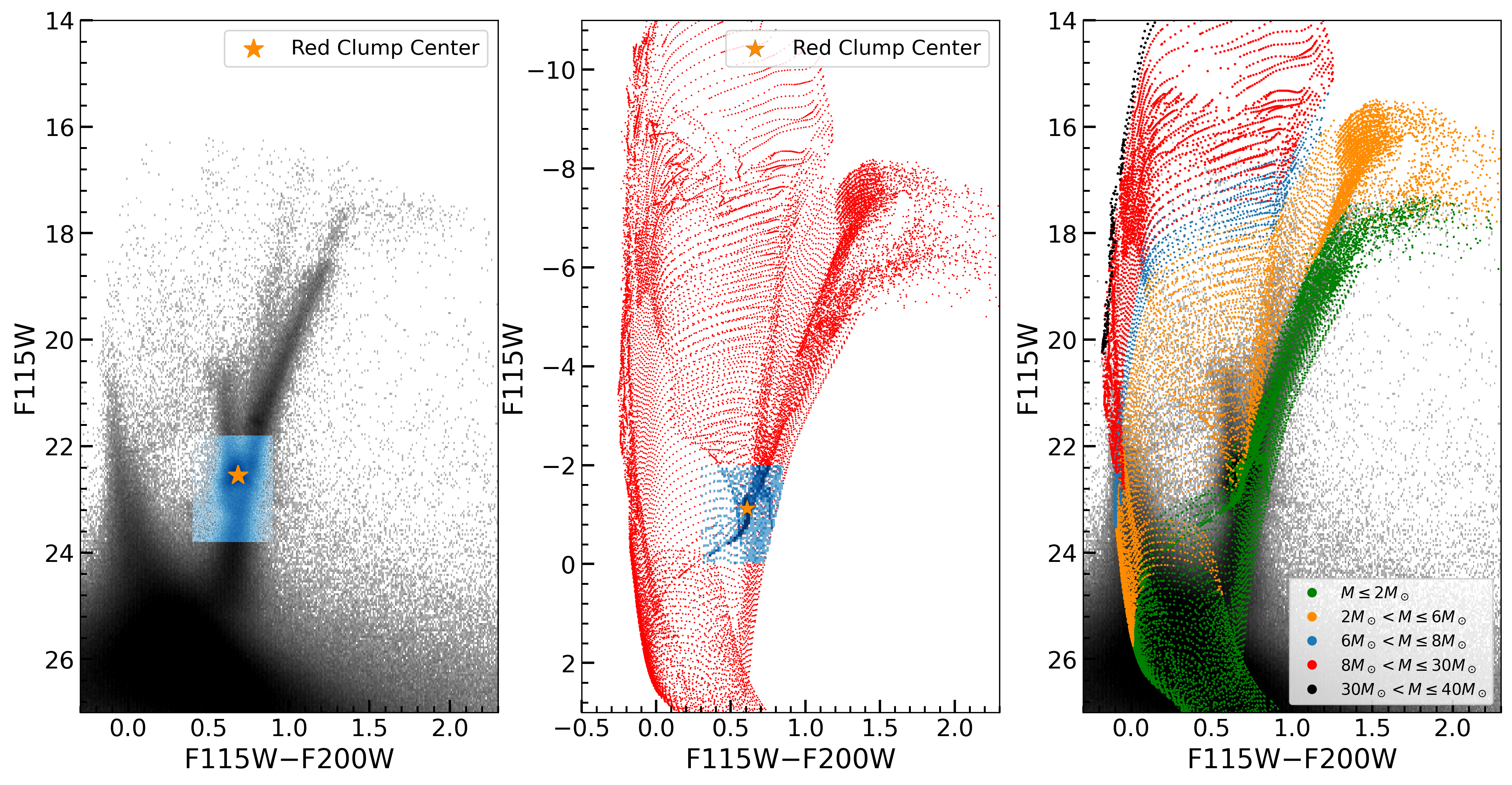}}
    \caption{Matching the observed CMD with the isochrones from the PARSEC model, shown here as an example for NGC 6822. The left and middle panels show the observed and model CMDs, respectively, while the right panel presents the matched CMD. In the left and middle panels, the blue region indicates the area used to calculate the maximum density point of the red clump, which is marked by the orange star. In the right panel, the green, orange, blue, red and black dots represent the PARSEC isochrones with stellar masses of $\leq 2~M_\odot$, $2$--$6~M_\odot$, $6$--$8~M_\odot$, $8$--$30~M_\odot$, and $30$--$40~M_\odot$, respectively. }
    \label{Figure.Parsec}
\end{figure}

In order to match the observed CMD with the PARSEC isochrones, the maximum-density point of the red clump is used as the reference point for alignment. First, the extinction of the observed CMD is corrected using $E(B-V) = 0.235$ \citep{1998ApJ...500..525S}, with $A_{F115W}/A_V = 0.319$ and $A_{F200W}/A_V = 0.133$, and assuming $R_V = 3.1$, as default in the CMD 3.8 web interface~\citep{1989ApJ...345..245C, 1994ApJ...422..158O}. The maximum-density positions of the red clump in both the observed CMD and the model CMD are then calculated, as represented by the orange star in Figure~\ref{Figure.Parsec}. The PARSEC isochrones are subsequently shifted by 0.073 mag to the right in color and by 23.675 mag downward in magnitude, according to the difference in the maximum-density positions. The aligned observed CMD and model are shown in the right panel of Figure~\ref{Figure.Parsec}.

With the aligned PARSEC isochrones, the regions of RSGs, O-AGBs, C-AGBs, and the Tip-RGB (TRGB) are defined on the observed CMD based on their theoretical loci. As shown in Figure~\ref{Figure.Cat_CMD}, the TRGB magnitude of each galaxy is indicated by the horizontal red dashed line, identified by visual inspection due to its prominent structure. The RSG region is defined by sources between $l_1$ and $l_2$ and brighter than the TRGB, the O-AGB region by sources between $l_2$ and $l_3$ and brighter than the TRGB, and the C-AGB region by sources located to the right of $l_3$ and brighter than the TRGB. The $l_1$ is defined as a linear function, and $l_2$ and $l_3$ are obtained by horizontally shifting $l_1$ by eye check, where the parameters of $l_1$, $l_2$, $l_3$ and TRGB magnitude are listed in Table~\ref{Table.CMD_boundary}. These regions provide a useful framework for distinguishing RSGs from foreground dwarfs on the CCD in the following analysis.

\subsection{Foreground Dwarfs on the CMD}
\begin{figure}
    \centerline{\includegraphics[width=0.35\linewidth]{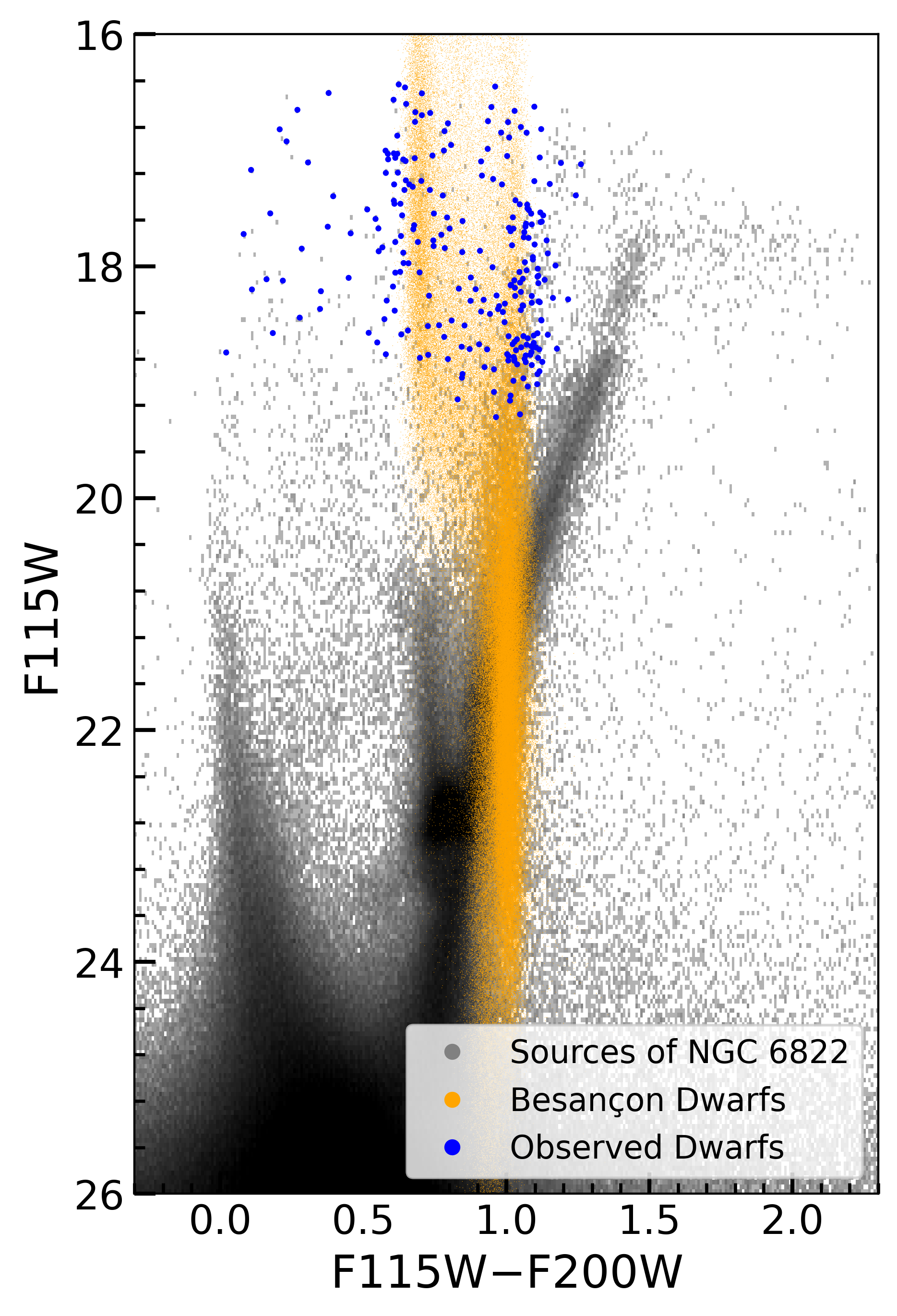}}
    \caption{Foreground dwarfs on the JWST CMD, shown here for NGC 6822 as an example. The black dots represent the observed data, the yellow dots represent foreground dwarfs from the Besan\c{c}on model, and the blue dots represent observed foreground dwarfs in \citet{2025ApJ...979..208L}.}
    \label{Figure.Besancon}
\end{figure}

The RSG region are contaminated by foreground dwarfs on the CMD, as they share similar apparent colors and magnitudes \citep{2021ApJ...923..232R, 2025ApJ...979..208L}.
After defining the RSG region on the JWST CMD, we examine the overlap between RSGs and foreground dwarfs on this CMD using the Besan\c{c}on model of Galactic stellar population synthesis~\citep{2003A&A...409..523R}.
Since the Besan\c{c}on model provides photometry only in the Johnson $J$, $H$, and $K$ bands, we first convert these to the JWST bands. By crossmatching the JWST sources with the confirmed foreground  dwarfs with the UKIRT data from \citet{2025ApJ...979..208L}, conversion coefficients are derived by fitting linear relations between F115W and $J$ band, and between F115W $-$ F200W and $J-K$ color, yielding $\text{F115W} = 0.950J + 1.124$ and $\text{F115W} - \text{F200W} = 0.736(J-K) + 0.388$.
As shown in Figure~\ref{Figure.Besancon}, the orange dots represent foreground dwarfs from the model after the magnitude and color conversions, which agree well with the observed foreground dwarfs (blue dots).
Both the Besan\c{c}on model and the observations demonstrate that foreground dwarfs overlap with the preliminary RSG region on the CMD.
Therefore, removing foreground dwarfs using the CCD method prior to identifying RSGs is necessary.

\subsection{Separation of RSGs and Foreground Dwarfs on the CCD}
The previously used $r-z/z-H$ and $J-H/H-K$ diagrams fail to distinguish RSGs from foreground dwarfs in metal-poor galaxies, because RSGs become bluer and blend into the foreground dwarf branch on these CCD as metallicity decreases, as observed in the SMC and LMC \citep{2025ApJ...979..208L}. In the other hand, most Local Group galaxies are dwarf irregular systems that are typically metal-poor, including those analyzed in this work (except NGC 300). Hence, developing a new CCD that works in metal-poor environment is important.
To find out the optimal CCD, we systematically construct multiple CCDs by combining the JWST F090W, F115W, F200W, F356W (or F335M), and F444W (or F430M) bands in all reasonable pairings. We then examine the distributions of foreground dwarfs and the RSGs identified by \citet{2025ApJ...979..208L} on each CCD.
By exploring these JWST color combinations, we find that F444W serves as an effective diagnostic band for separating RSGs from foreground dwarfs, even in metal-poor environments. 
Among all combinations, the most effective color combination is F115W $-$ F200W versus F356W $-$ F444W, which shows a clear separation between foreground dwarfs and member RSGs.

\begin{figure}
    \centerline{\includegraphics[width=1\linewidth]{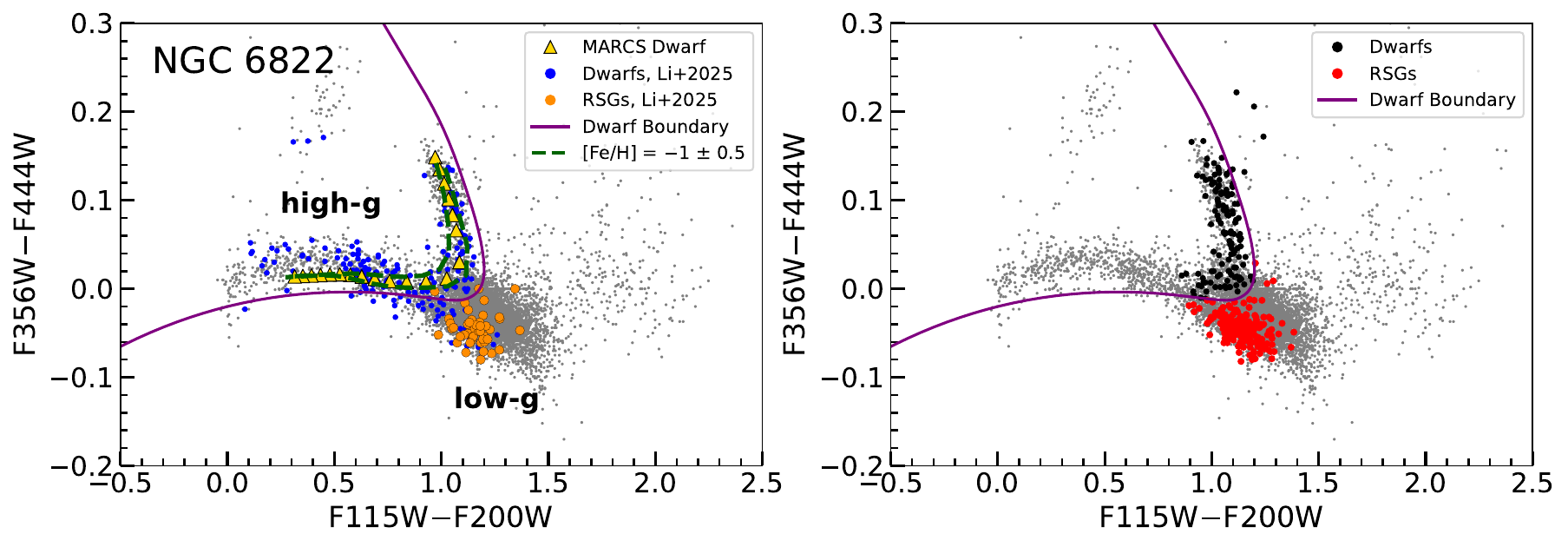}}
    \caption{The optimal CCD for metal-poor environments, F115W $-$ F200W versus F356W $-$ F444W, shown here for NGC 6822 as an example. 
    In both panels, the gray dots represent all analyzed sources, and the solid line indicates the adopted boundary of the dwarf branch.
    In the left panel, the blue and orange dots indicate the removed foreground dwarfs and identified RSGs by \citet{2025ApJ...979..208L}, respectively, the yellow triangles show the dwarf tracks from the MARCS stellar atmosphere model,
    and the green dashed lines represent the dwarf tracks from the MARCS model within [Fe/H]$=-1.0\pm 0.5$.}
    In the right panel, the red dots denote the RSG candidates identified on the CCD in this work, while the black dots represent the removed foreground dwarfs. 
    
    \label{Figure.CCD_NGC6822}
\end{figure}

We use both observational data and MARCS stellar atmosphere model to demonstrate the distribution of foreground dwarfs and RSGs on this CCD.
As shown in the left panel of Figure~\ref{Figure.CCD_NGC6822}, a clear separation between the removed foreground dwarfs (blue dots) and identified RSGs (orange dots) by \citet{2025ApJ...979..208L} is evident, which is not revealed in previously used CCDs.
The MARCS stellar atmosphere model is then used to confirm the dwarf branch. The input parameters of the MARCS model are as follows. A standard chemical composition and a plane-parallel geometry are adopted. The $T_\text{eff}$ ranges from 3500 to 4000 K in steps of 100 K and from 4000 to 8000 K in steps of 250 K. We adopt [Fe/H] = $-1.0$ to match the metallicity of stars at high Galactic latitudes in the Milky Way, $\log g$ = 4.5, and a microturbulent velocity of 1.0 km s$^{-1}$. The model spectra are convolved with the corresponding JWST filter response curves from the SVO\footnote{\url{http://svo2.cab.inta-csic.es/svo/theory/fps3/index.php?mode=browse}} filter library to derive synthetic magnitudes and colors. Reddening is then applied to the theoretical dwarf colors to match the observations, adopting $E(B-V) = 0.235$~\citep{1998ApJ...500..525S} for NGC 6822, using the same extinction coefficients mentioned above.
The dwarf tracks from the MARCS model are represented by yellow triangles in the left panel of Figure~\ref{Figure.CCD_NGC6822}, which align closely with the observed foreground dwarfs (blue dots).

\begin{figure}
    \centerline{\includegraphics[width=0.55\linewidth]{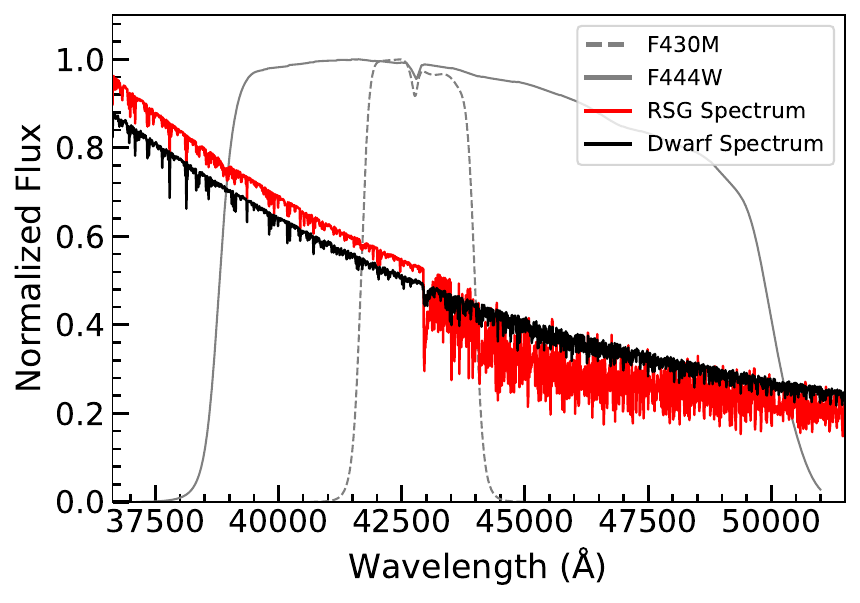}}
    \caption{Model spectra of a RSG (red line) and a dwarf (black line) from the MARCS stellar atmosphere model in the JWST F444W band, with the same $T_{\mathrm{eff}}$ = 4000 K at [Fe/H] = $-1.0$.
    The gray solid and dashed lines indicate the F444W and F430M filter responses, respectively. The strong CO absorption feature in the RSG spectrum reduces the flux in the F444W or F430M band.
    The fluxes of the RSG and dwarf spectra are normalized to the maximum RSG flux and scaled by the same factor for display.}
    \label{Figure.Spectrum_F444W}
\end{figure}

The key factor that produces this separation between RSGs and foreground dwarfs is the CO absorption bands in the F444W (or the F430M) band, which is prominent in RSG spectra. As shown in Figure~\ref{Figure.Spectrum_F444W}, the RSG spectrum (red line) exhibits strong CO absorption, while the dwarf spectrum (black line) does not. This absorption reduces the F356W $-$ F444W color of RSGs, causing them to shift downward on the CCD and become well separated from the foreground dwarfs. This property makes the JWST CCD a reliable tool for distinguishing RSGs from foreground dwarfs in metal-poor galaxies.

The boundary of the dwarf branch on the CCD is defined using the theoretical dwarf tracks from the MARCS model, and subsequently used to remove foreground dwarfs. We perform cubic spline interpolation on the theoretical tracks to derive the central locus of the dwarf branch. 
The MARCS models predict the intrinsic color locus of dwarf stars at a given metallicity and surface gravity. However, the observed dwarf colors exhibit significant dispersion due to variations in metallicity, surface gravity, and photometric uncertainties. As a result, dwarfs form a broadened branch rather than a thin theoretical line in the CCD.
Hence, this locus is then shifted by 0.019 mag downward and 0.120 mag to the right to construct the adopted boundary, corresponding to a dispersion of the dwarf branch of approximately $1.5\sigma$ in F356W $-$ F444W and $2.7\sigma$ in F115W $-$ F200W, which is computed using the confirmed foreground dwarfs by \citet{2025ApJ...979..208L}.
This dwarf boundary is shown as the solid line in Figure~\ref{Figure.CCD_NGC6822}. Sources to the left of the boundary are high log $g$ objects (foreground dwarfs), and sources to the right are low log $g$ objects (RSGs and AGBs). 
To address the effect of metallicity variations in the MARCS models on the dwarf boundary, we construct dwarf model tracks at $[\mathrm{Fe/H}] = -1.0 \pm 0.5$, as shown by the green dashed lines in Figure~\ref{Figure.CCD_NGC6822}, and use their envelope as a proxy for the confidence interval of the theoretical dwarf locus. All these dwarf tracks remain on the left side of the adopted boundary in the CCD, indicating that metallicity variations of foreground dwarfs within $-1.0\pm 0.5$ dex do not shift the dwarf locus into the RSG region. Therefore, this boundary can effectively remove dwarf contamination.
Some observed dwarfs from \citet{2025ApJ...979..208L} fall below the boundary, which is expected because they are faint sources with large photometric uncertainties.

\subsection{Removing the Foreground Dwarfs on the CCD}
By the clear separation between RSGs and foreground dwarfs on the F115W $-$ F200W versus F356W $-$ F444W diagram in metal-poor galaxies, foreground dwarfs can be directly removed using the defined boundary of the dwarf branch. In each galaxy, we analyze sources with $m < (m^{\rm TRGB} + 1.5)$ mag on the CMD, as including too many unrelated faint sources weakens the separation on the CCD. For NGC 6822, this magnitude-limited sample contains 8,091 sources, of which 7,900 have both SW- and LW-band photometry, while the remaining 191 are detected only in the SW bands.
For sources with both the SW- and LW-band photometry, we first select those falling within the RSG region on the CMD and then further removing the foreground dwarfs using the CCD. The sources detected only in the SW bands are also included to ensure the completeness of the RSG sample and are classified solely based on the CMD.

Among the 7,900 sources with both SW and LW photometry, 330 sources fall within the defined RSG region on the CMD and are subsequently decontaminated on the CCD. As shown in the right panel of Figure~\ref{Figure.CCD_NGC6822}, 127 sources located to the left of the dwarf boundary (solid line) are removed as foreground dwarfs. In addition, five sources located in the upper-right region of the CCD are also removed, because they are too far from the expected RSG location to be genuine RSGs and they are verified as redder foreground dwarfs on other CCDs. A total of 132 sources are removed as foreground dwarfs (black dots). The remaining 198 sources located to the right of the dwarf boundary are classified as uncontaminated RSG candidates (red dots), denoted as the RSG-CCD sample.
The 191 sources detected only in the SW bands are classified on the CMD. Among them, 10 sources fall within the RSG region and are included as a supplementary subset, denoted as the RSG-CMD sample.
Based on the fraction of foreground dwarfs removed in the CCD selection, the contamination rate of the RSG-CMD sample is estimated to be approximately 40\%, implying that about four out of the ten sources may not be genuine RSGs.

In total, 208 RSG candidates are identified in NGC 6822, consisting of 198 RSG-CCD sample and 10 RSG-CMD sample. The summary of the sample selection and source statistics for NGC 6822 is listed in Table~\ref{Table.star_number}.

\section{Application to Other Galaxies}
\label{Sect.Application}
\subsection{Sextans A and NGC 300}
\label{Sect.SexA_NGC300}
\begin{figure}
    \centerline{\includegraphics[width=1\linewidth]{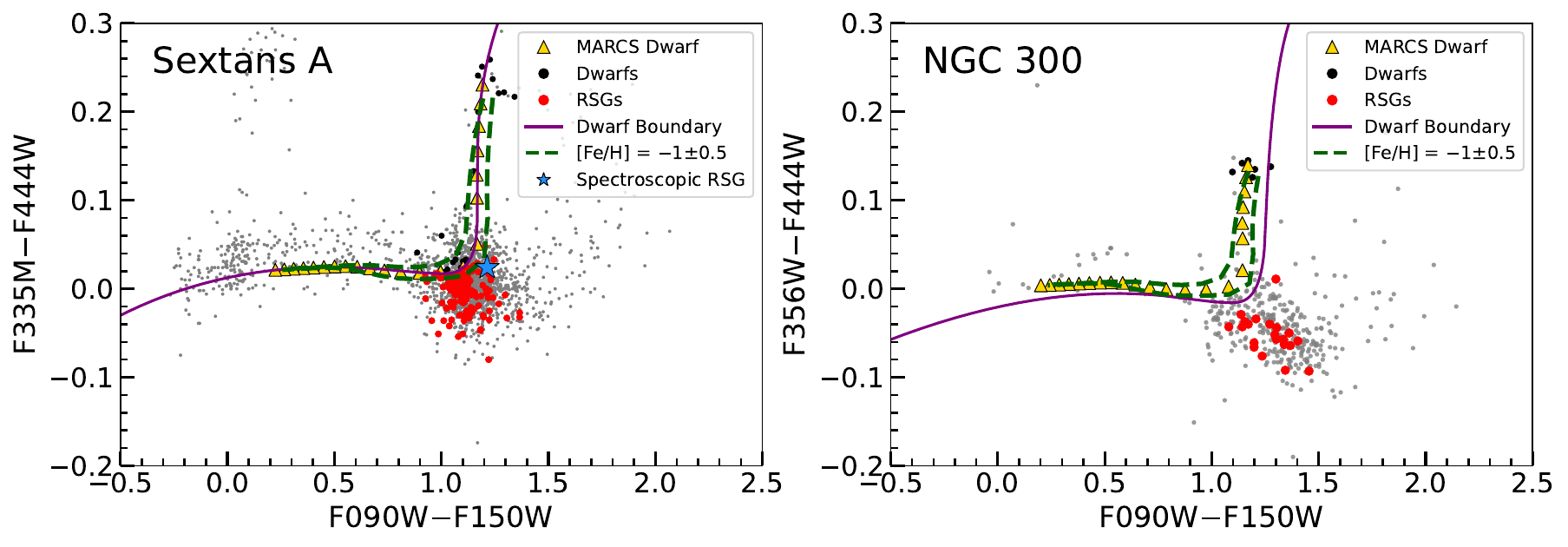}}
    \caption{Alternative CCDs constructed using the F444W band for Sextans A and NGC 300. 
    The symbols are the same as those in Figure~\ref{Figure.CCD_NGC6822}.
    The blue star in the left panel represents the spectroscopically confirmed RSG reported by \citet{2019A&A...631A..95B}.}
    
    \label{Figure.CCD_SexA_NGC300}
\end{figure}

Sextans A and NGC 300 are analyzed following the same procedure as NGC 6822 to obtain the pure RSG samples. However, due to the absence of certain filters required to construct the optimal F115W $-$ F200W versus F356W $-$ F444W diagram, alternative color combinations are adopted: F090W $-$ F150W versus F335M $-$ F444W for Sextans~A and F090W $-$ F150W versus F356W $-$ F444W for NGC 300, as shown in Figure~\ref{Figure.CCD_SexA_NGC300}. The reddening values applied to the MARCS dwarf models are $E(B-V)=0.044$ for Sextans A and $E(B-V)=0.013$ for NGC 300~\citep{1998ApJ...500..525S}, using the same extinction coefficients as described above.

\begin{figure}
    \centerline{\includegraphics[width=1.0\linewidth]{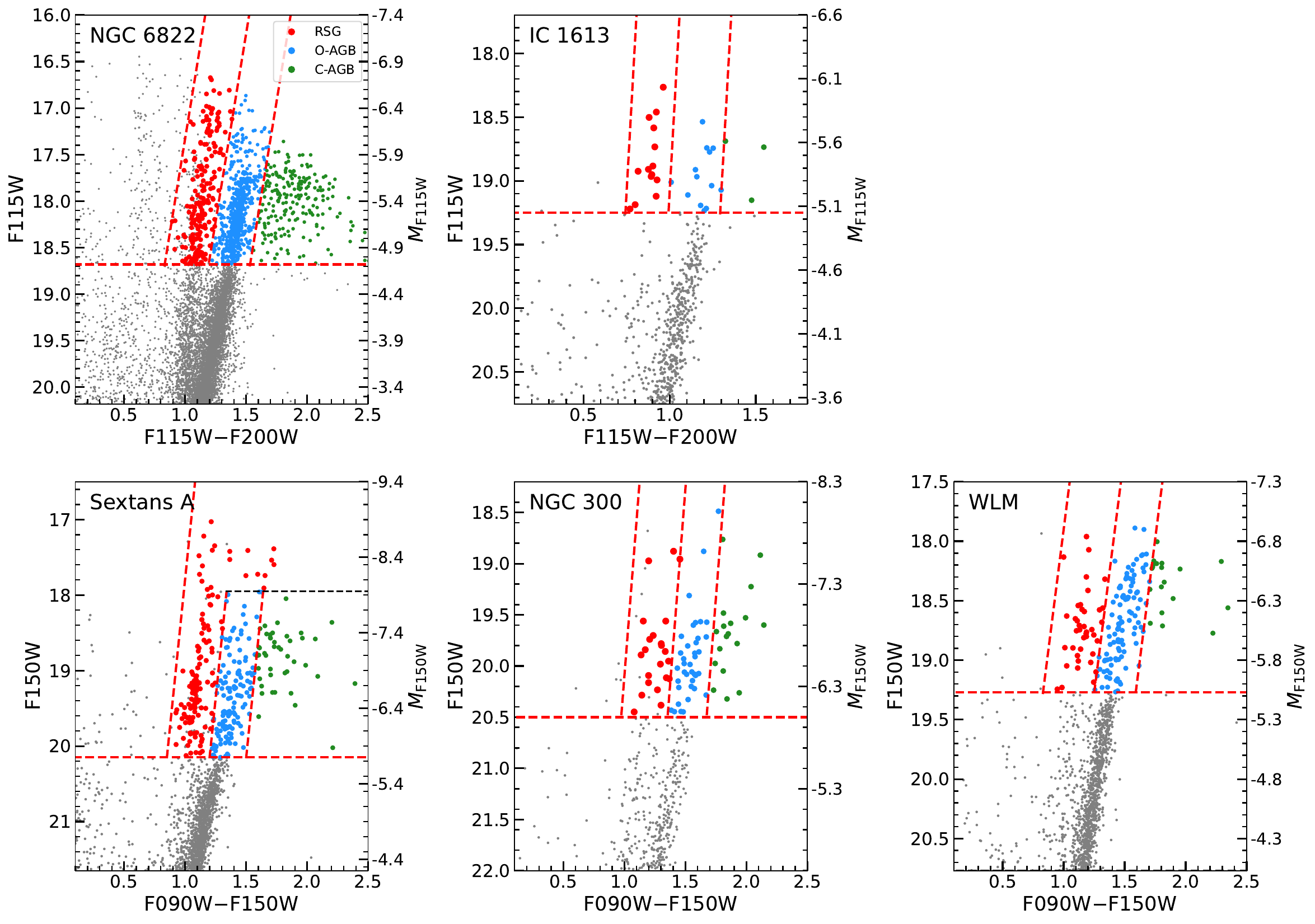}}
    \caption{The identified RSG (red dots), O-AGB (blue dots), and C-AGB (green dots) candidates in NGC 6822, IC 1613, Sextans A, NGC 300, and WLM on the CMD. The gray dots represent all analyzed sources. The adopted distance moduli are $m - M = 23.40$ for NGC~6822 \citep{2012MNRAS.421.2998F}, $26.52$ for NGC~300 \citep{2009ApJS..183...67D}, and $24.25$, $25.95$, and $24.77$ for IC~1613, Sextans~A, and WLM~\citep{2012AJ....144....4M}, respectively, to plot absolute magnitudes on the right-hand axis.}
    \label{Figure.5CMD}
\end{figure}

For Sextans A, the dwarf boundary is defined by the central locus of the MARCS dwarfs as shown in the left panel of Figure~\ref{Figure.CCD_SexA_NGC300}, since the RSGs on this CCD are strongly concentrated. Moreover, the small field of view of Sextans A minimizes foreground contamination. As a result, among the 149 sources within the RSG region on the CMD that have both SW- and LW-band data, 127 sources located to the right of the dwarf boundary are selected as RSG candidates for the RSG-CCD sample, while 22 sources are removed as foreground dwarfs (17 lying to the left of the dwarf boundary and five lying in the upper-right region of the CCD). 
Specifically, one spectroscopically confirmed RSG (hereafter sRSG, named Sextans A9 in their work) reported by \citet{2019A&A...631A..95B} is crossmatched with this RSG-CCD sample using a search radius of $0.4^{\prime\prime}$.
This sRSG is shown by the blue star in the left panel of Figure~\ref{Figure.CCD_SexA_NGC300}. sRSG is the key anchor of the RSG sample. This identical source makes a nice illustration of the selection method and makes this photometric sample reliable.
There are no sources that fall within the RSG region on the CMD among the objects detected only in the SW bands in Sextans A.
Besides, a specific RSG-CMD sample with high mass-loss rates is further defined for Sextans A. As shown in the Sextans A panel of Figure~\ref{Figure.5CMD}, sources above the black horizontal dashed line are classified as RSGs with high mass-loss rates. This line corresponds to apparent and absolute magnitudes of 17.95 and $-8.00$ mag, respectively, assuming a distance modulus of $m - M = 25.95$ \citep{2012AJ....144....4M}. This selection is motivated by the fact that these sources are separated from AGB stars by their distinct CMD locations, and that RSGs can reach bolometric absolute magnitudes as bright as $M_{\mathrm{bol}} \sim -9$ mag. Similar classification criteria have been adopted in \citet{2019A&A...629A..91Y,2021A&A...646A.141Y}, \citet{2020ApJ...900..118N}, and \citet{ 2021AJ....161...79M}. Eight sources are classified as high mass-loss RSG candidates in Sextans A, which are added to the RSG-CMD sample. In total, 135 RSG candidates are identified in Sextans A.

For NGC~300, the MARCS dwarf locus is shifted by 0.100 mag to the right and by 0.012 mag downward to define the dwarf boundary, so as to include foreground dwarfs, as shown in the right panel of Figure~\ref{Figure.CCD_SexA_NGC300}.
Because the positional variation of foreground dwarfs in the CCD remains small, the shifts for NGC 300 are defined using those of NGC 6822 as a reference, but allowing for modest adjustments. These differences arise from variations in Galactic foreground metallicity along different lines of sight, differences in photometric dispersion, and uncertainties associated with extinction correction and reddening.
As a result, 21 out of 29 sources located to the right of the boundary are identified as RSG candidates for the RSG-CCD sample, while 8 sources are removed as foreground dwarfs (five lying to the left of the dwarf boundary and three lying in the upper-right region of the CCD). In addition, one source is included in the RSG-CMD sample. In total, 22 RSG candidates are identified in NGC 300.

The summary of sample selection and source statistics for Sextans A and NGC 300 is listed in Table~\ref{Table.star_number}.

\subsection{IC 1613 and WLM}
For IC 1613, the RSG candidates are directly identified on the CMD because the F444W band is unavailable. A total of 14 RSG candidates are identified in IC 1613 and are denoted as the RSG-CMD sample.

For WLM, although the F430M band is used as a substitute for F444W, the CCD based on F090W, F150W, and F430M is less effective than the $\mathrm{color}_{\mathrm{SW}}$ versus $\mathrm{color}_{\mathrm{LW}}$ diagram in distinguishing RSGs from foreground dwarfs. Even though color combinations involving $\sim 4~\mu$m bands generally provide better separation than previously used colors, the F356W $-$ F444W color yields the most effective separation. Therefore, the RSG candidates in WLM are identified on the CMD. 
In total, 40 RSG candidates are identified in WLM and are denoted as the RSG-CMD sample.

Fortunately, foreground contamination is negligible owing to the small field of view of these two galaxies. 
The summary of sample selection and source statistics for IC 1613 and WLM is listed in Table~\ref{Table.star_number}.

\section{Sample of RSG and AGB Candidates}\label{Sect.results}

\subsection{RSG candidates}
\begin{figure}
    \centerline{\includegraphics[width=0.9\linewidth]{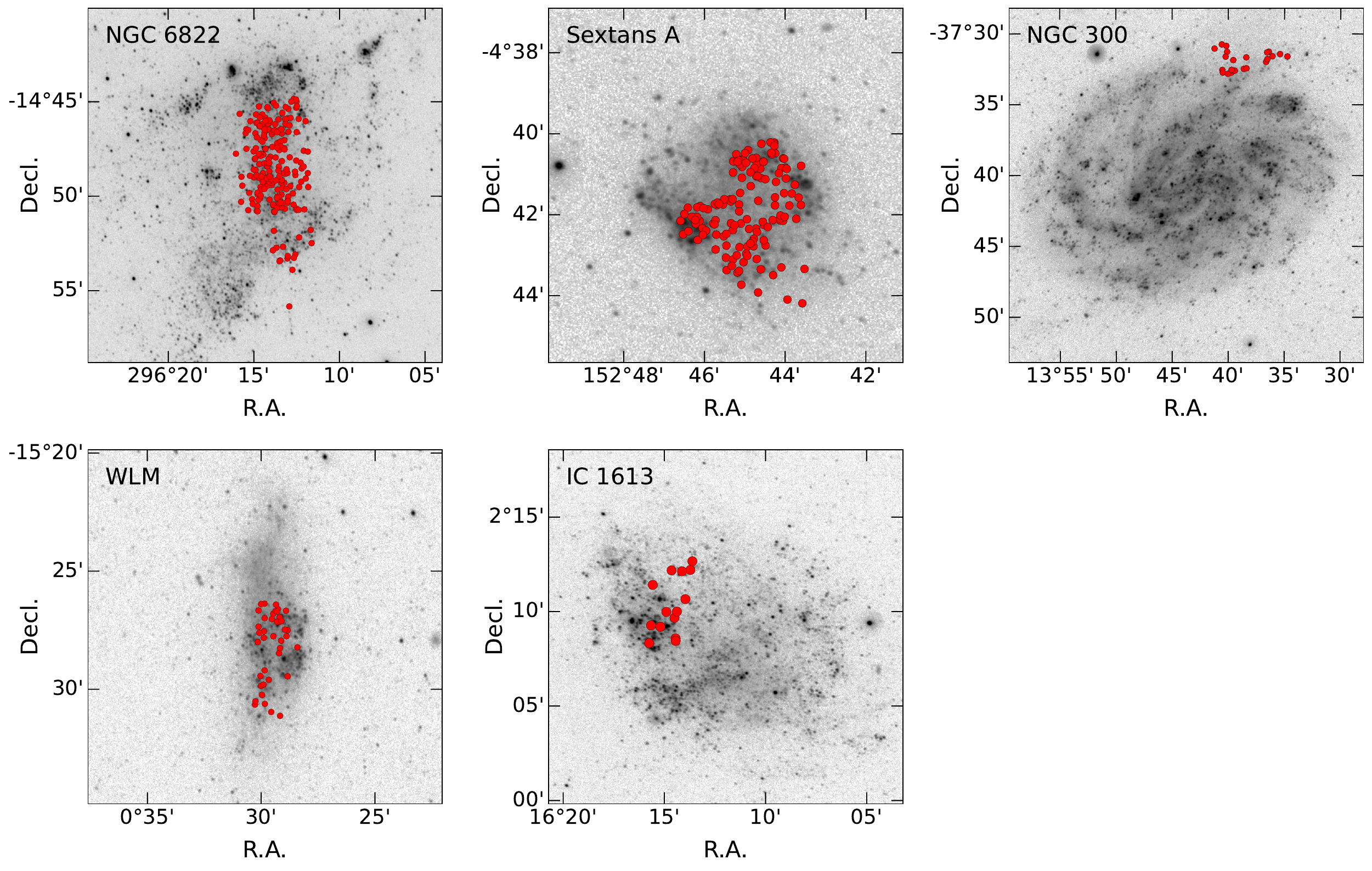}}
    \caption{The RSG candidates overlaid on the GALEX ultraviolet images, highly overlapped with the star-forming regions.}
    \label{Figure.Overlap_RSG}
\end{figure}

The distributions of RSG candidates in the five galaxies on the CMDs are displayed in Figure~\ref{Figure.5CMD}, where the RSGs are represented by red dots.
In total, 208 RSG candidates are identified in NGC 6822 (198 and 10 from the RSG-CCD and RSG-CMD samples, respectively), 135 in Sextans~A (127 and 8 from the RSG-CCD and RSG-CMD samples, respectively), 22 in NGC 300 (21 and 1 from the RSG-CCD and RSG-CMD samples, respectively), 40 in WLM and 14 in IC 1613 (RSG-CMD sample). 
The RSG candidate catalog is presented in Table~\ref{Table.RSG_AGB_cat} and is available in machine-readable form.
The RSG candidates overlaid on the GALEX images are shown in Figure~\ref{Figure.Overlap_RSG}, highly consistent with the star-forming regions of these galaxies.

\subsection{O-AGB and C-AGB candidates}
The sample O-AGB and C-AGB stars are the by-products of this work, identified directly from the CMD, as there are no foreground dwarfs red enough to contaminate these regions. Their identification is valuable for studying AGB stellar populations, constraining dust production in metal-poor galaxies~\citep{2009A&A...506.1277G, 2024ApJS..275...33W}, and refining galaxy distance measurements~\citep{2024ApJ...967...22L}.
Sources located within the O-AGB and C-AGB regions on the CMD are classified accordingly. In total, 583 O-AGB and 216 C-AGB candidates are identified in NGC 6822; 136 O-AGB and 51 C-AGB candidates in Sextans A, 38 O-AGB and 18 C-AGB candidates in NGC 300, 107 O-AGB and 23 C-AGB candidates in WLM, and 13 O-AGB and 3 C-AGB candidates in IC 1613. 
The AGB candidate catalog is also included in Table~\ref{Table.RSG_AGB_cat} and is available in machine-readable form.
Their distributions on the CMD are shown in Figure~\ref{Figure.5CMD}, where the O-AGB candidates are indicated by blue dots and the C-AGB candidates by green dots.

\section{Comparison with previous works}\label{Sect.comparison}
Here we compare our results with previous studies. Only seven RSG candidates are reported in Sextans A by \citet{2019A&A...631A..95B}. For WLM, seven are reported by \citet{2019A&A...631A..95B}, and eleven by \citet{2006ApJ...648.1007B} and \citet{2012AJ....144....2L}. For IC 1613, three are identified by \citet{2019A&A...631A..95B}, eight by \citet{2019MNRAS.483.4751H}, and 14 by \citet{2022ApJ...939...28C}. In NGC 300, 27 bright RSG candidates are reported by \citet{2015ApJ...805..182G}. Most of these RSG candidates are identified through spectroscopic observations and evolutionary model analyses. In the work of \citet{2021ApJ...923..232R}, the number of RSG candidates increases to 33 in Sextans A, 63 in WLM, and 115 in IC 1613, using the $J-H/H-K$ and $r-z/z-H$ diagrams to remove foreground dwarfs. This represents the largest catalog of RSG candidates in these galaxies to date.

In NGC 6822, \citet{2019A&A...629A..91Y} identify 234 RSG candidates using the $r-z/z-H$ diagram to remove foreground dwarfs. Later, \citet{2020ApJ...892...91H} identify 1,292 candidates by selecting the intersection of three CMDs, although their sample suffers from a contamination rate of 52.93\%. \citet{2021ApJ...923..232R} identify 465 RSG candidates by combining the $J-H/H-K$ and $r-z/z-H$ diagrams to remove foreground dwarfs. \citet{2022AJ....163...70D} report 51 candidates using Gaia astrometric data. A more complete and purer sample is later presented by \citet{2025ApJ...979..208L}, who combine the CCD and Gaia astrometric methods to remove foreground dwarfs and further trace the RSG region to recover potential RSGs blended into the dwarf branch on the CCD. In their study, 1,184 and 843 RSG candidates are identified in the complete and pure samples, with contamination rates from foreground dwarfs of 20.5\% and 6.5\%, respectively. We further exclude 299 O-AGB contaminants from their pure sample using the $r-z$ versus $z$ diagram, as suggested by their study, resulting in a final sample of 544 RSG candidates for comparison.

Considering the completeness and purity of previously identified samples, we compare our RSG candidates in Sextans A, WLM, and IC 1613 with those from \citet{2021ApJ...923..232R}, and those in NGC 6822 with the samples from \citet{2025ApJ...979..208L}.
For NGC 300, since the RSG candidates reported by \citet{2015ApJ...805..182G} are confined to the bright end which are not included in our sample, no comparison is made.

\begin{figure}
    \centerline{\includegraphics[width=0.85\linewidth]{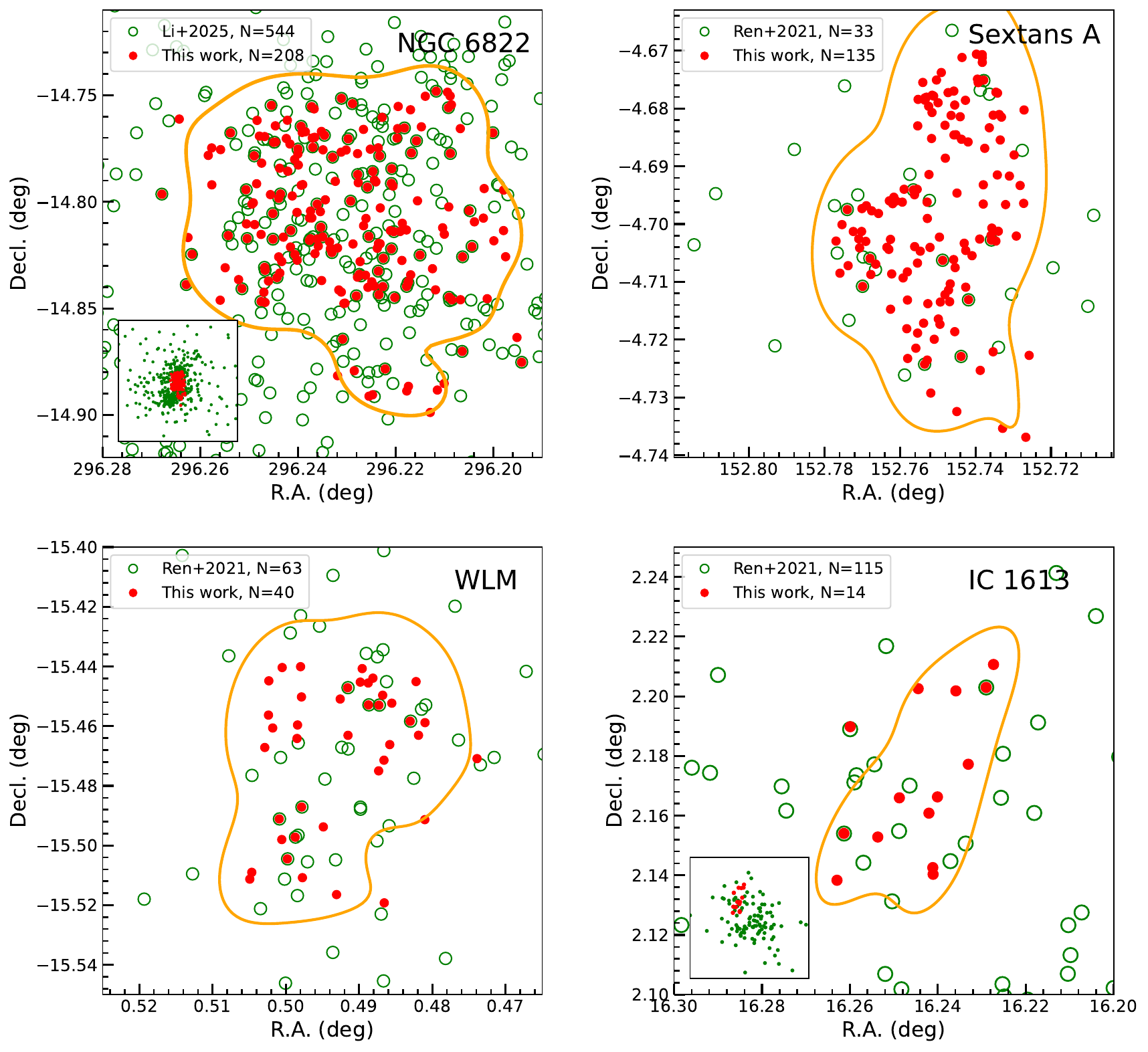}}
    \caption{The spatial distribution of RSG candidates in comparison with \citet{2021ApJ...923..232R} for Sextans A, WLM, and IC 1613, and with \citet{2025ApJ...979..208L} for NGC 6822. In each panel, red dots and green circles represent the RSG candidates from this work and from previous studies, respectively. The yellow line shows the boundary of the JWST observation region, obtained by contouring the RSG candidates in this work, where the marginal density of the contour is 15\% of the maximum.
    In the NGC 6822 and IC 1613 panels, the lower-left inset shows the full RSG distribution of previous studies.}
    \label{Figure.Comparison}
\end{figure}

The spatial distributions of RSG candidates from this work (red dots) and previous studies (green circles) are shown in Figure~\ref{Figure.Comparison}. The number of RSG candidates in Sextans A and in the central region of NGC 6822 is noticeably increased compared to previous studies. In Sextans~A, the sample increases from 40 in \citet{2021ApJ...923..232R} to 135 in this study. The samples from \citet{2021ApJ...923..232R} and \citet{2025ApJ...979..208L} cover larger sky areas and wider magnitude ranges ($\sim4$--$5$~mag) for RSGs, as they are based on the ground UKIRT observations, which provide wide-area coverage and appropriate exposure times to include the bright stars. In contrast, the JWST RSG sample spans a narrower magnitude range ($\sim1$--$2$~mag) due to saturation at the bright end, and JWST's smaller field of view further limits coverage. Therefore, standardizing the comparison is necessary.
To standardize the magnitude range, we first crossmatch our sources with those from previous studies. We then consider only sources that are fainter than the brightest source in this cross-match in each sample.
To unify the sky coverage, we define a boundary (the yellow line in Figure~\ref{Figure.Comparison}) based on our sample by contouring, where the marginal density of the contour is 15\% of the maximum, and only sources within this boundary are included. 
This contour defines the same sky coverage as adopted in this work, enabling a consistent comparison with previous studies, which cover larger sky areas. After applying the same spatial constraints, the comparison clearly demonstrates the improvement in RSG identification enabled by JWST.
The choice of the 15\% marginal density level best matches the spatial distribution of the RSGs identified in this work, closely tracing the main RSG concentration without inappropriately enlarging or shrinking the effective region. We test several thresholds (5\%, 10\%, 15\%, 20\%, and 30\%). Lower thresholds (e.g., 5\%–10\%) produce overly extended contours that include sparsely populated outer regions, while higher thresholds (e.g., 20\%, 30\%) generate contours that are too restrictive and exclude a significant fraction of RSGs. Using contouring or the footprint to define the boundary has a negligible impact on the comparison. After applying these two constraints, the previous studies yielded 112, 17, 21, and 4 RSG candidates for NGC 6822, Sextans A, WLM, and IC 1613, respectively. These numbers are small compared with the 208, 135, 40, and 14 RSG candidates identified in this work. Among these, 61, 9, 8, and 2 sources are common to both studies. For sources identified only in previous works, some are classified as O-AGB candidates in this study (e.g., 14 sources in NGC 6822) owing to the improved photometric precision and spatial resolution. The remaining unmatched sources are likely due to the slightly different borderlines of the defined RSG region on the CMD or differences in data quality control.

Spectroscopically confirmed RSGs (sRSGs) are an important anchor of the RSG sample. Therefore, we also perform crossmatching with previously reported sRSG samples.
\citet{2015ApJ...803...14P} report 18 sRSGs in NGC 6822. \citet{2019A&A...631A..95B} report 7 sRSGs in Sextans A, 4 in WLM, and 3 in IC 1613. \citet{2024A&A...686A..77B} report 46 sRSGs in NGC 300. \citet{2025A&A...698A.279D} report 10 sRSGs in NGC 6822 and 2 in IC 1613.
Among all of these sRSG samples, we identify only one common source in Sextans A reported by \citet{2019A&A...631A..95B}, as described in Section~\ref{Sect.SexA_NGC300}.
After detailed checking, the unmatched sources are because their confirmed sRSGs are generally too bright for the JWST observations or there is no spatial overlap.

These results demonstrate the significant improvement enabled by JWST observations, which allow a more complete and reliable census of RSGs. However, current JWST observations of Local Group galaxies (LGGs) are limited by the small field of view, as illustrated by the observation footprints in Figure~\ref{Figure.Observation}, with only one or two JWST pointings available for each galaxy. In addition, many existing JWST observations adopt long exposure times to reach deep photometric limits and construct full CMDs of LGGs, which often results in saturation at the bright end, leading to the loss of massive stars such as RSGs and Wolf-Rayet stars. As a consequence, constructing a complete RSG sample currently requires combining JWST data with observations from ground-based telescopes.
Nevertheless, JWST remains highly suitable for studies of massive stars in LGGs. Given the relatively small angular sizes and moderate distances of most LGGs, three to five JWST pointings with shorter exposure times would be sufficient to cover their main bodies. A dedicated JWST survey of LGGs with wider spatial coverage and optimized exposure times would therefore be highly valuable for probing the bright end of the stellar populations in these galaxies.

\section{Summary}\label{Sect.conclusion}
In this study, we perform PSF photometry on publicly released JWST observations for five Local Group galaxies: NGC~6822, Sextans~A, NGC~300, WLM, and IC~1613, using the DOLPHOT NIRCam module and conduct a systematic identification of RSG populations based on these data.
The unprecedented spatial resolution, photometric precision, and depth of JWST enable a significant improvement over previous ground-based studies, particularly in crowded and metal-poor environments where RSG identification has long been challenging.

RSGs and foreground dwarfs overlap on the CMD as they share similar apparent colors and magnitudes.
By testing multiple NIRCam filter combinations, we find an optimized color-color diagram (CCD) for metal-poor environments, i.e., F115W $-$ F200W versus F356W $-$ F444W. This CCD provides a clear separation between RSGs and foreground dwarfs and effectively mitigates contamination from foreground dwarfs that has limited earlier optical and near-infrared CCD approaches.
The boundary of the foreground dwarf branch on the CCD is determined by fitting and shifting theoretical dwarf tracks from the MARCS model.
RSG candidates are first selected on the CMD within the predefined RSG region and are then further decontaminated on the CCD for NGC~6822, Sextans~A, and NGC~300. For WLM and IC~1613, RSG candidates are identified directly on the CMD because the effective CCD is unavailable.
This procedure provides a robust framework for identifying RSGs with high reliability across different Local Group galaxies.

Applying this approach, we identify 208, 135, 22, 40, and 14 RSG candidates in NGC~6822, Sextans~A, NGC~300, WLM, and IC~1613, respectively. 
Compared with previous studies covering the same luminosity range and sky region, the number of identified RSG candidates increases significantly. 
In particular, the number of RSG candidates in Sextans~A increases from 40 in earlier works to 135 in this study, and a clear enhancement is also observed in the central region of NGC~6822.
As a by-product, we also identify samples of O-AGB and C-AGB candidates in these galaxies.

Despite these advances, current JWST observations of Local Group galaxies remain limited by small fields of view and saturation at the bright end. Future JWST programs with more pointings and optimized exposure times will be essential for conducting a complete census of massive stars. In forthcoming work, we will extend this methodology to an expanded sample of approximately ten additional Local Group galaxies as soon as new JWST data become available, enabling a comprehensive and homogeneous study of RSG populations across a wide range of metallicities and star-forming environments. 

\section*{Acknowledgements}
This work is supported by the National Key R\&D Program of China through grant No.\ 2025YFF0511004, National Natural Science Foundation of China through grant Nos.\ 12133002, 12373048, and 12203025, China Manned Space Project through grant No.\ CMS-CSST-2025-A14, Shandong Provincial Natural Science Foundation through project ZR2022QA064 and Shandong Provincial University Youth Innovation and Technology Support Program through grant No.\ 2022KJ138. This work has made use of the data from JWST.

\software{astropy \citep{2013A&A...558A..33A,2018AJ....156..123A},
          TOPCAT \citep{2005ASPC..347...29T}, dustmaps \citep{2018JOSS....3..695G}, dolphot \citep{2000PASP..112.1383D, 2016ascl.soft08013D, 2024ApJS..271...47W}}

\bibliography{sample631}{}
\bibliographystyle{aasjournal}

\begin{table*}
\centering
\caption{Summary of JWST Observations Used in This Work}
\label{Table.jwst_observation}
\begin{tabular}{llcccl}
\toprule
Galaxy & Date & Program ID & Observtn & Exposure Time (s) & Bands \\
\midrule
NGC 6822 & 2022-09-05 & 1234 & 6, 10 & $4 \times 150$ & F115W, F200W, F356W, F444W \\
NGC 6822 & 2022-09-15 & 1234 & 9     & $4 \times 128$ & F115W, F200W, F356W, F444W \\
Sextans A & 2023-04-15 & 1619 & 14, 15 & $4 \times 311$ & F090W, F150W, F335M, F444W \\
NGC 300  & 2022-10-16 & 1638 & 2 & $4 \times 257$ & F090W, F150W, F356W, F444W \\
WLM      & 2022-07-23, 24  & 1334 & 5      & $4 \times 7537$ & F090W, F430M \\
WLM      & 2022-07-24  & 1334 & 5      & $4 \times 5862$ & F150W \\
IC 1613  & 2023-08-18 & 2391 & 6 & $4 \times 1030$ & F115W, F150W \\
IC 1613  & 2023-08-18 & 2391 & 6 & $4 \times 2104$ & F200W, F335M \\
\bottomrule
\end{tabular}
\end{table*}

\begin{table*}[htbp]
\centering
\caption{Summary of Photometry Results}
\label{Table.photometry_result}
%\footnotesize
\fontsize{8.}{9.5}\selectfont
\begin{tabular}{lccccc}
\toprule
Galaxy & SW & SW+LW & Crossmatched (Search radius) & QC & Final Data \\
\midrule
NGC 6822  & 4,698,860 & 4,420,299 & 3,625,521 ($0.06^{\prime\prime}$) & 754,084 & 724,604   \\
Sextans A & 910,714 & 860,771 & 615,192 ($0.05^{\prime\prime}$)       & 152,384 & 144,028  \\
NGC 300   & 352,082  & 343,226  & 198,949 ($0.05^{\prime\prime}$)    & 31,073  & 31,071    \\
WLM      & 2,153,335 & 2,063,601 & 1,787,127 ($0.05^{\prime\prime}$) & 495,781 & 495,405   \\
IC 1613  & 1,348,770 & 1,301,291 & 1,062,043 ($0.05^{\prime\prime}$) & 222,719 & 222,699   \\
\bottomrule
\end{tabular}
\begin{minipage}{0.95\textwidth}
\footnotesize
\tablecomments{ SW and LW denote the short-wavelength and long-wavelength of NIRCam, respectively.
QC refers to quality-controlled sources, selected using the photometric flags of F115W and F200W for NGC~6822 and IC~1613,
and F090W and F150W for Sextans~A, NGC~300, and WLM.
The final dataset is obtained after removing duplicated sources from the QC sample.}
\end{minipage}
\end{table*}

\begin{table*}[htbp]
\centering
\caption{JWST/NIRCam Photometric Catalog for the Five Local Group Galaxies}
\label{Table.photometry_catalog}
\fontsize{8.}{9.5}\selectfont
\setlength{\tabcolsep}{4pt}
\begin{tabular}{lccccccccccc}
\toprule
Galaxy & RA (deg) &Dec (deg) & F115W & Err\_F115W &
F200W & Err\_F200W &F356W &Err\_F356W &F444W &$\cdots$ &
Err\_F430M \\
\midrule
NGC~6822 & 296.244390 & $-14.846028$ & 17.397 & 0.001 & 15.996 & 0.001 & 15.806 & 0.001 & 15.867 & $\cdots$ &  \\
NGC~6822 & 296.270360 & $-14.782633$ & 17.637 & 0.002 & 15.935 & 0.001 & 15.215 & 0.001 & 15.176 & $\cdots$ &  \\
NGC~6822 & 296.234509 & $-14.773234$ & 18.387 & 0.002 & 15.782 & 0.001 & 99.999 & 9.999 & 99.999 & $\cdots$ &  \\
NGC~6822 & 296.234882 & $-14.836645$ & 17.607 & 0.001 & 15.926 & 0.001 & 15.367 & 0.001 & 15.380 & $\cdots$ &  \\
NGC~6822 & 296.264384 & $-14.771735$ & 17.757 & 0.001 & 15.880 & 0.001 & 15.080 & 0.001 & 15.000 & $\cdots$ &  \\
\ldots   & \ldots     & \ldots        & \ldots & \ldots & \ldots & \ldots & \ldots & \ldots & \ldots & \ldots & \ldots \\
\bottomrule
\end{tabular}
\vspace{2mm}
\begin{minipage}{0.95\textwidth}
\footnotesize
\tablecomments{ This table combines the photometric catalogs of all five galaxies into a single dataset, which can be  filtered by galaxy name.
The available NIRCam bands for each galaxy (F090W, F115W, F150W, F200W, F335M, F356W, F430M, and F444W) 
are listed in Table~\ref{Table.jwst_observation}.
RA and Decl are in degrees (J2000).
Magnitudes are in the VEGAMAG system.
Blank entries indicate that the corresponding band is not included for that galaxy.
The values 99.999 (magnitudes) and 9.999 (magnitude uncertainties) represent numeric null values.}
The full catalog is available in machine-readable form.

\end{minipage}
\end{table*}

\FloatBarrier

\begin{table*}[htbp]
\centering
\caption{Boundaries of RSGs and AGBs on the CMD in Figure~\ref{Figure.Cat_CMD} }
\label{Table.CMD_boundary}
\begin{tabular}{lccccc}
\hline
\hline
Galaxy & $k$ & $b$ & Shift$_1$ & Shift$_2$ & TRGB Mag (band) \\
\hline
NGC 6822 & $-8.07$ & $25.40$ & $0.36$ & $0.70$ & $18.67$ (F115W) \\
IC 1613  & $-24.14$ & $37.20$ & $0.25$ & $0.55$ & $19.25$ (F115W)\\
Sextans A & $-15.81$ & $33.60$ & $0.35$ & $0.65$ & $20.15$ (F150W)\\
NGC 300  & $-15.25$ & $35.35$ & $0.38$ & $0.70$ & $20.50$ (F150W)\\
WLM      & $-8.10$ & $26.00$ & $0.42$ & $0.76$ & $19.27$ (F150W)\\
\hline
\end{tabular}
\begin{minipage}{0.95\textwidth}
\footnotesize
\tablecomments{ $l_1$ is defined as a linear function, where $k$ and $b$ represent the slope and intercept of $l_1$, respectively. The boundaries of $l_2$ and $l_3$ are obtained by horizontally shifting $l_1$ by Shift$_1$ and Shift$_2$, respectively.}
\end{minipage}
\end{table*}

\begin{table*}[htbp]
%\centering
\caption{Summary of Sample Selection and Source Statistics}
\label{Table.star_number}
\begin{tabular}{lccccccc}
\toprule
Galaxy & SW+LW & SW-only & RSG-CCD & RSG-CMD & Total RSG & O-AGB & C-AGB \\
\midrule
NGC 6822  & 7900  & 191  & 198 &  10   & 208 & 583 & 216 \\
Sextans A & 1933  &  51  & 127 &  8   & 135 & 136 &  51 \\
NGC 300  & 412    &  30  &  21 &  1   &  22 &  38 &  18 \\
WLM      & \dots & 1412  & \dots & 40  & 40 &  107 &  23 \\
IC 1613  & \dots &  606  & \dots &  14  &  14 &  13 &   3 \\
\bottomrule
\end{tabular}
\end{table*}

\begin{table*}
\centering
\caption{Identified RSG and AGB Candidates Catalog for the Five Galaxies}
\label{Table.RSG_AGB_cat}
\fontsize{8.}{9.5}\selectfont
\setlength{\tabcolsep}{4pt}
\begin{tabular}{lccccccccccc}
\toprule
Galaxy & RA (deg) & Dec (deg) &F115W & Err\_F115W &F200W & Err\_F200W &
F356W & Err\_F356W &F444W  &$\cdots$ & StellarClass \\
\midrule
NGC~6822 & 296.267871 & $-14.796390$ & 17.294 & 0.001 & 16.091 & 0.001 & 15.864 & 0.001 & 15.917 & $\cdots$ & RSG-CCD \\
NGC~6822 & 296.232367 & $-14.790379$ & 17.389 & 0.002 & 16.214 & 0.001 & 16.088 & 0.002 & 16.109 & $\cdots$ & RSG-CCD \\
NGC~6822 & 296.248474 & $-14.791508$ & 17.472 & 0.001 & 16.226 & 0.001 & 15.998 & 0.001 & 16.066 & $\cdots$ & RSG-CCD \\
NGC~6822 & 296.257647 & $-14.792015$ & 17.415 & 0.001 & 16.172 & 0.001 & 15.994 & 0.001 & 16.026 & $\cdots$ & RSG-CCD \\
NGC~6822 & 296.241535 & $-14.821190$ & 17.410 & 0.001 & 16.162 & 0.001 & 16.020 & 0.001 & 16.083 & $\cdots$ & RSG-CCD \\
\ldots   & \ldots     & \ldots       & \ldots &\ldots & \ldots &\ldots &\ldots  &\ldots & \ldots & \ldots   & \ldots \\
\bottomrule
\end{tabular}
\vspace{2mm}
\begin{minipage}{0.95\textwidth}
\footnotesize
\tablecomments{
This table follows the same format as Table~\ref{Table.photometry_catalog}, 
but includes only the identified RSG and AGB candidates.
The final column StellarClass is added to indicate the stellar classification based on the selection criteria.
The full catalog is available in machine-readable form.
}
\end{minipage}
\end{table*}

\FloatBarrier

\end{CJK*}
\end{document}